\documentclass[pra,a4paper,aps,onecolumn,showpacs,superscriptaddress,groupedaddress]{revtex4}
\usepackage{ae}
\usepackage[T1]{fontenc}
\usepackage[ansinew]{inputenc}
\usepackage{amsmath}
\usepackage{amssymb}
\usepackage{graphicx}
\usepackage{color}
\usepackage[colorlinks]{hyperref}
\usepackage{lscape}
\begin{document}
\title{Correlations In $n$-local Scenario}
\author{Kaushiki Mukherjee}
\email{kaushiki_mukherjee@rediffmail.com}
\affiliation{Department of Applied Mathematics, University of Calcutta, 92, A.P.C. Road, Kolkata-700009, India.}
\author{Biswajit Paul}
\email{biswajitpaul4@gmail.com}
\affiliation{Department of Mathematics, St.Thomas' College of Engineering and Technology, 4, Diamond Harbour Road, Alipore, Kolkata-700023, India.}
\author{Debasis Sarkar}
\email{dsappmath@caluniv.ac.in, debasis1x@yahoo.co.in}
\affiliation{Department of Applied Mathematics, University of Calcutta, 92, A.P.C. Road, Kolkata-700009, India.}

\begin{abstract}
Recently Bell-type inequalities were  introduced in Phys. Rev. A \textbf{85}, 032119 (2012) to analyze the correlations emerging in an entanglement swapping scenario characterized by independence of the two sources shared between three parties. The corresponding scenario was referred to as \textit{bilocal} scenario. Here, we derive Bell-type inequalities in $n+1$ party scenario, i.e., in $n$-local scenario. Considering the two different cases with several number of inputs and outputs, we derive local and $n$-local bounds. The $n$-local inequality studied for two cases are proved to be tight. Replacing the sources by maximally entangled states for two binary inputs and two binary outputs and also for the fixed input and four outputs, we observe quantum violations of $n$-local bounds. But the resistance offered to noise cannot be increased as compared to the bilocal scenario. Thus increasing the number of parties in a linear fashion in source independent scenario does not contribute in lowering down the requirements of revealing quantumness in a network in contrast to the star configuration (Phys. Rev. A \textbf{90}, 062109 (2014)) of $n+1$ parties.
\end{abstract}
\date{\today}
\maketitle

\section{Introduction}

Correlation is one of the most important word with the study of foundational aspects of quantum mechanics. By correlations we basically focus on the relation between the outputs of the measurements performed on composite quantum systems. Recently, the study of correlations is gaining importance to construct the theoretical background of many computational tasks \cite{Acin,Mayer,Pironio,Colbeck,Bancal}. Specifically, in quantum key distributions nonlocal correlations play an important role and it enables us to understand the behavior of nonlocal correlations in much more profound way. Apart from its fundamental interest, the study of nonlocal correlations is important for several other aspects of quantum information theory \cite{Cleve,A. Acin,Briegel,Ra}. To detect the nonlocal nature of quantum systems, Bell inequalities (or, Bell-type inequalities) \cite{bell1} play a major role and also provide us with criteria suitable for categorizing correlations.
In a system of three parties namely Alice, Bob and Charlie the correlations compatible with a local causal model (\cite{Bel}) can be written in the form:
\begin{equation}\label{t}
P(a,b,c|x,y,z)=\int d\lambda \rho(\lambda)P(a|x,\lambda)\times P(b|y,\lambda)\times P(c|z,\lambda)
\end{equation}
where $x$, $y$ and $z$ represent the inputs of Alice($A$), Bob($B$) and Charlie($C$) respectively and $a$, $b$ and $c$ are their outputs,  $\lambda$ is the joint hidden state following the distribution  $\rho(\lambda)$ and satisfying the normalization condition:  $\int d\lambda \rho(\lambda)$=1. The correlations which cannot be written in this form (\ref{t}) are said to be nonlocal.
\begin{figure}[htb]
\centering
\includegraphics[width=5in]{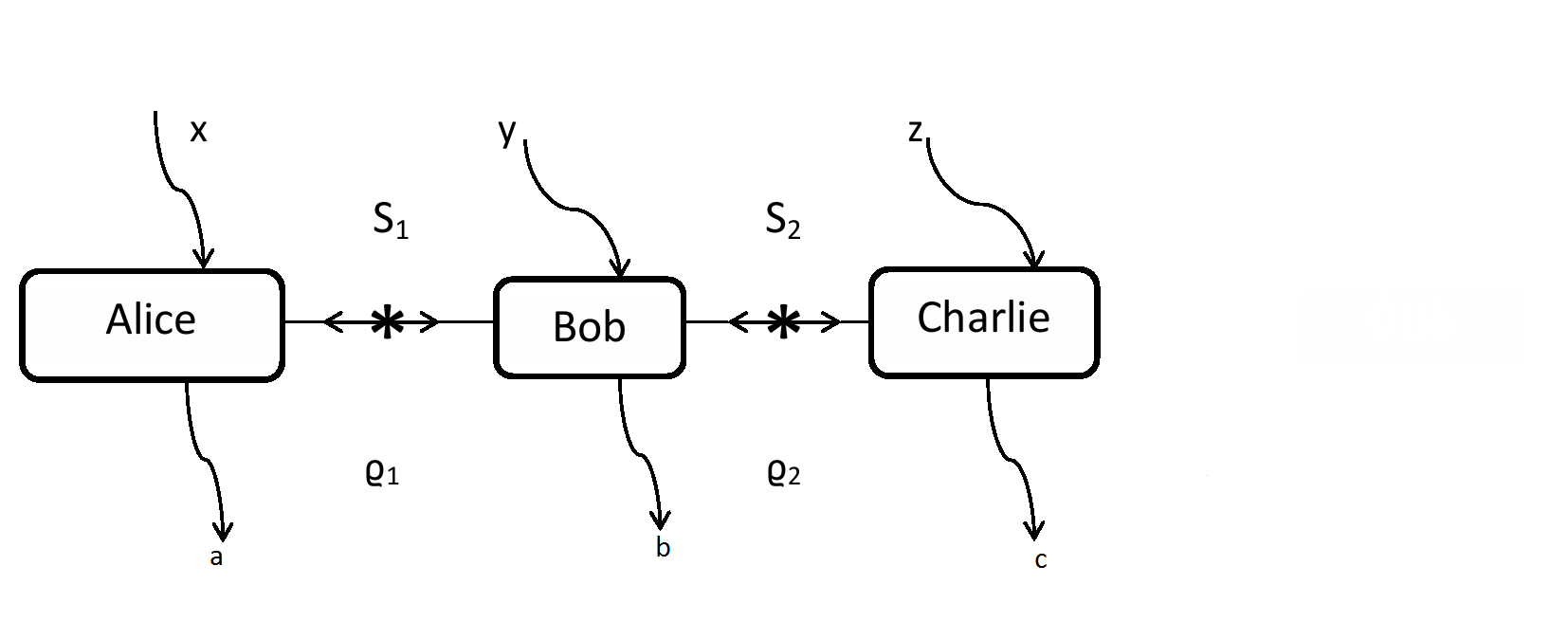}
\caption{\emph{Entanglement swapping scenario where each of $3$ parties Alice, Bob and Charlie share  sources $S_1$ and $S_2$ (each emit independent pairs of particles in some quantum states $\rho_1$ and $\rho_2$ respectively). The intermediate party Bob will perform a joint measurement on their particles that it receives from sources $S_{1}$ and $S_2$. The final state is an entangled state shared between Alice and Charlie who then apply local measurements $x$ and $z$ on their particles and obtain outcomes $a$ and  $c$ respectively. This type of experiment is usually
characterized by a joint probability distribution $P(a,b,c|x,y,z)$. }}
\end{figure}
\\
 The study of correlations  between the results of measurements performed in quantum networks has recently gained much interests. In some future quantum networks, like in \cite{Briegel} and \cite{HAM}, a process known as entanglement swapping \cite{Zukowski} is used. This is a process by which particles that never interacted directly can also become correlated nonlocally(see FIG. 1.). To analyze and characterize nonlocal properties of correlations generated in a network it is interesting to  assume source independence in the network, i.e., to consider models where the independent systems are characterized by uncorrelated hidden states($\lambda_i$). In \cite{BRA}, a theoretical framework was introduced to address broadly the role of nonlocality in entanglement swapping contexts and shown that this additional assumption of source independence leads to stronger tests of nonlocality. They considered a  three party scenario where the sources shared by the parties are assumed to be independent of each other. Such a model was referred to as `\textit{Bilocal scenario}' and the corresponding correlations as `\textit{bilocal correlations}'. In a bilocal scenario(see FIG. 2.), there are two sources $S_1$ and $S_2$ shared  between three parties Alice($A$), Bob($B$) and Charlie($C$) arranged in a linear way such that any two neighboring parties share a common source. For the intermediate party B, who received two particles, the measurement will typically be a joint measurement on both the particles received by him. Under this, joint probability distribution is defined as,
\begin{equation}\label{E}
P(a, b, c|x, y, z)=\iint d\lambda_1d\lambda_2\rho_1(\lambda_1)\rho_2(\lambda_2)P(a|x, \lambda_1)P(b|y, \lambda_1, \lambda_2)P(c|z, \lambda_2)
\end{equation}
where $\lambda_1$ characterizes the joint hidden state of the system produced by the source $S_1$ and $\lambda_2$ is for the system $S_2$. The hidden states $\lambda_1, \lambda_2$ follow independent probability distributions $\rho_1(\lambda_1)$ and $\rho_{2}(\lambda_2)$ such that
\begin{equation}\label{F}
    \int\rho_1(\lambda_1)d \lambda_1=\int\rho_2(\lambda_2)d \lambda_2=1.
\end{equation}
The definition (\ref{E}) actually follows from the Bell's locality assumption (\ref{t}), just by considering one extra assumption
\begin{equation}\label{G}
    \rho(\lambda_1, \lambda_2)=\rho_1(\lambda_1)\rho_2(\lambda_2).
\end{equation}
\begin{figure}[htb]
\centering
\includegraphics[width=5in]{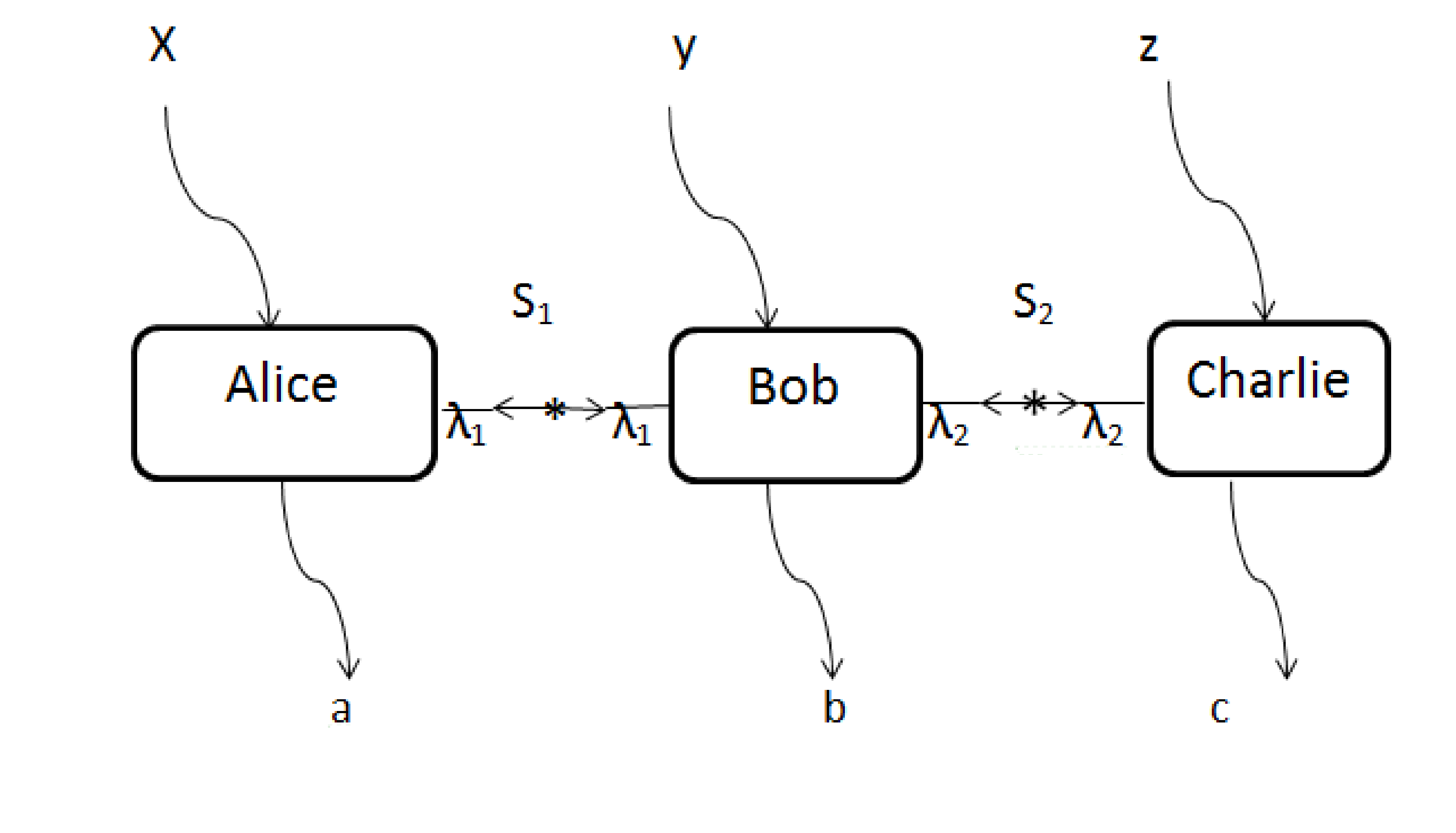}
\caption{\emph{ Bilocal scenario where three parties Alice, Bob and Charlie share two sources $S_1$ and $S_2$. The source $S_1$ sends particles to Alice and Bob and source $S_2$ sends particles to Bob and Charlie. The sources $S_1$ and $S_2$ are characterized by the hidden states $\lambda_1$ and $\lambda_2$ respectively.  All parties can perform measurements on their systems,  labeled by $x$,  $y$ and $z$  for Alice,  Bob and Charlie and  they obtain outcomes denoted by $a$,  $b$  and $ c$  respectively. Bob  might perform a joint measurement on the two particles that he receives from $S_1$ and $S_2$. The sources are assumed to be independent}.}
\end{figure}\\
 In recent times, source independence has been explored to enrich study of correlations \cite{BRAN,Fri,Frit,Hen,Tav}. For instance, this assumption of independence of sources is found to be important to study detection loophole in some local models \cite{Gisin,Greenberger}. Besides, in \cite{BRAN} source independence($bilocal$) assumption was exploited to increase the resistance offered to noise by states used in a $bilocal$ network compared to the resistance offered by a state in a standard $CHSH$ scenario \cite{Cla}. This in turn lowers the level of restrictions to be imposed on experiments demonstrating quantumness in a network (e.g., entanglement swapping).  Hence the study of bilocal correlations can be applied in various fields of quantum computation such as device independent information processing (\cite{Acin,Mayer}), private randomness generation (\cite{Pironio, Colbeck}), device independent entanglement witnesses (\cite{Bancal}), etc. From this perspective, apart from linear arrangement of parties and sources, in \cite{Fri} various other `correlation scenarios' characterized by source independence were studied  where each of the parties involved in the scenario was supposed to perform a single measurement.  In particular in \cite{Tav} A. Tavakoli et. al. dealt with a star configuration of parties where they showed that the resistance to noise will increase further if the number of parties is increased in a non-linear pattern. In this context, one may ask whether resistance in a source independent scenario can be increased by increasing number of parties in a linear pattern. In this paper, we focus on this question, however arriving to the intuition that unlike non-linear pattern, generalization of the $bilocal$ scenario to a linear $n$-local scenario is of no use in this regard. For that we have studied correlations in $n+1$ party system characterized by source independence and hence exploited the $n$-local scenario(which will be discussed in Section II). In particular we have considered two scenarios differing on the basis of number of inputs and outputs of intermediate parties compatible with various experiments. In course of work we have given $n-$local and local Bell-type inequalities along with instances of non $n-$local but local quantum correlations which in turn exploits the utility of source independence for demonstrating quantumness in a more natural way compared to standard nonlocal(Bell-CHSH scenario) in a network involving $n+1$ parties.   \\

In short, the paper is organized as follows: in section II, we introduce the concept of $n$-locality. We derive non-linear $n$-local and local Bell-type inequalities in two different scenarios in section(III). In section(IV) we check whether the quantum correlations produced in entanglement swapping scenario (VA) and by partial Bell-state measurement (VB) violate $n$-local inequality or not. Finally, we will present our conclusion in section (V).

Throughout this paper,  we deal all the cases with finite number of possible inputs and outputs.

\section{$n$-local Scenario}

\subsection{Basic Assumptions for $n$-local Scenario}
\begin{figure}[htb]
\centering
\includegraphics[width=5in]{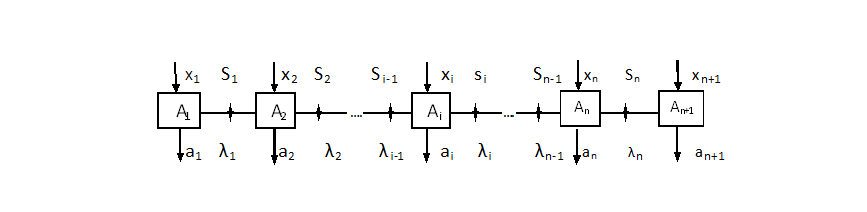}
\caption{\emph{The general scenario where each of $n+1$ parties $A_i$ share $n$ sources $S_i(i=1,...,n)$. $S_i$ sends particles to $A_i$ and $A_{i+1}$. Each source is characterized by the hidden state $\lambda_i$. Each party $A_i$ can perform measurement on their systems, labeled by $x_i$ and the outcomes denoted by $a_i$. Each of the $n-1$ intermediate parties $A_{i}(i=2,...,n)$ will perform a joint measurement on their particles. The sources are assumed to be  independent}.}
\end{figure}
The $n$-local scenario is depicted in FIG. 3. There are $n$ sources $S_i(i=1,...,n)$ and $n+1$ parties $A_i(i=1,...,n+1)$ arranged in a linear way such that any two neighboring parties share a common source. For each $i(i=1,...,n+1)$, party $A_i$ can perform two dichotomic measurements $x_i=x_i^k(k=1,2)$ where $x_i^k\in\{0,1\}$ on the system they have received and obtain outcomes $a_i=a_i^j(j=1,2)$ where $a_i^j\in\{0,1\}$. Excepting the extreme two parties $A_1$ and $A_{n+1}$ each of the remaining $n-1$ parties who receives two subsystems, the measurement will be a joint measurement operating on both subsystems simultaneously. In this scenario, Bell's locality assumption takes the form:
\begin{equation}\label{K}
    P(a_1,....,a_{n+1}|x_1,...,x_{n+1})=\int d\lambda \rho(\lambda)\Pi_{i=1}^{n+1}P(a_i|x_i, \lambda).
\end{equation}
Here $\lambda$ is the joint hidden state. For each party $A_i(i=1,...,n+1)$ the input is $x_i$ and the corresponding output $a_i$, is determined by the local distributions $P(a_i|x_i, \lambda)$. The hidden state $\lambda$ follows the distribution  $\rho(\lambda)$,  satisfying the normalization condition $\int d\lambda \rho(\lambda)=1$. Here we also assume that the measurement choices of each party are independent of $\lambda$.\\

Now let each source $S_i$ be characterized by hidden state $\lambda_i$. Moreover we assume that for each $i\in\{2,...,n\}$ the outputs of party $A_i$ depend on the states $\lambda_{i-1}$, $\lambda_{i}$ and its input $x_i$ whereas for $A_1$ and $A_{n+1}$ the outputs depend on their corresponding inputs and also $\lambda_1$ and $\lambda_n$ respectively, i.e., for each party its output is dependent on the states that it receives from the adjacent sources and on the type of measurements performed on those systems,  but not on the measurements performed on the distant systems, i.e.,  we can write,
\begin{equation}\label{L}
P(a_1,...,a_{n+1}|x_1,...,x_{n+1})= \int d\lambda_{1}...\int d\lambda_n \rho(\lambda_{1},..., \lambda_{n})P(a_1|x_1, \lambda_{1})\Pi_{i=2}^{n}P(a_i|x_i, \lambda_{i-1},\lambda_{i})P(a_{n+1}|x_{n+1}, \lambda_{n}).
\end{equation}
Without any further assumption, Eq.(\ref{L}) is equivalent to Eq.(\ref{K}). In particular, $\rho(\lambda_{1},..., \lambda_{n})$ is different from zero only when  the hidden states are the same, i.e., $\lambda_{i}=\lambda_{i+1}(i=1,...,n-1)$ to recover Eq.(\ref{K}).
Now we make the $n$-local assumption: Under this assumption of source independence, the distribution of the hidden states $\lambda_{i}(i=1,n)$ will be factorized as below,
\begin{equation}\label{M}
\rho(\lambda_{1},..., \lambda_{n})=\Pi_{i=1}^n\rho_i(\lambda_{i}).
\end{equation}
 As $n$ sources are supposed to be independent,  we assume that the property characterized by the equation Eq.(\ref{M}) carries over to the local model(characterized by the equation Eq.(\ref{K})).
\\
 Eq.(\ref{L}) and Eq.(\ref{M}) together define the assumptions on $n$-locality. Each of the hidden states $\lambda_{i}$ now follows an independent distribution $\rho_i(\lambda_{i})$ such that  $\int d\lambda_{i} \rho_i(\lambda_{i})=1\,\forall i=1,...,n$. Other than the fact that each $\lambda_{i}$ is measurable,  no further restriction is made on the domain of these variables (as in the case of bilocality).\\
For $n=3$ the scenario introduced here is different from that discussed in \cite{BRAN} where a four partite experiment under restrictions of source independence and fixed joint measurement settings was compared with a standard Bell scenario. The correlations obtained therein were thus similar to that one can obtain in a standard Bell experiment between two parties. But the three-local scenario framed in this paper is a four party experiment performed under the assumption of source independence and also that of free will, i.e., each of the four parties can choose freely between two dichotomic measurement settings. This approach is also different from that of Fritz in \cite{Fri} where he considers one measurement setting per party. Our work is mainly motivated to extend the idea of source independence in a $n$ party scenario arranged in linear pattern keeping the assumption of free will intact and thereby develop the corresponding Bell-type inequalities.
\subsection{Topological features of the $n$-local set}
The topology of the $n$-local set is the same as that of the \textit{bilocal} one. Topological features of the $n$-local set are thus summed up below:\\
\begin{center}
\begin{enumerate}
  \item $n$-local correlation being local (by construction),  the set of  $n$-local correlations $\mathcal{L}$ is a  subset  of the local set ($\mathcal{T}$): $\mathcal{L}\subseteq\mathcal{T}$
  \item $\mathcal{L}$ is not convex, as mixture of $n$-local correlations is not necessarily $n$-local (due to nonlinearity constraint (\ref{M})).
  \item Being extremal points of the local set,  deterministic correlations are also $n$-local, i.e., $\mathcal{L}$  is the convex hull of $\mathcal{T}$.
\end{enumerate}
\end{center}

\subsection{Representation of the $n$-local correlation in terms of $q_{\bar{\alpha_1}\bar{ \alpha_2} \bar{\alpha_3}\bar{\alpha_4}}$}

If a local correlation is written in the form (\ref{K}), then we know that each party's (say $A_1$) local response function $P(a_1|x_1, \lambda_1)$ can be taken to be deterministic,  i.e., for each input $x_1$ exactly one output $a_1$ would be obtained. In case of finite number of such deterministic strategies corresponding to an assignment of an output $\alpha_1^{x_1}$ to each of $A_1$'s $N$ possible inputs $x_1^k(k=1,...,N)$, we denote each of these strategies of party $A_1$  by the string $\bar{\alpha_{1}}=\alpha_{11}\alpha_{21}...\alpha_{N1}$ and also denote the corresponding response function by $P_{\bar{\alpha_1}}(a_1|x_1)=\delta_{a_1,\alpha_{1}^{x_1}}$. The corresponding response function for the remaining $n$ parties can be defined in a similar pattern. Using this,  $n$-local correlations can be alternatively defined. We could write equation (\ref{L}) in an equivalent form with $q_{\bar{\alpha_1}...,\bar{ \alpha_{n+1}}}$ as
\begin{equation}\label{N}
   P(a_1,...,a_{n+1}|x_1,...,x_{n+1}) = \sum_{\bar{\alpha_1},...,\bar{ \alpha_{n+1}}} q_{\bar{\alpha_1}...\bar{ \alpha_{n+1}}}\Pi_{i=1}^{n+1}P_{\bar{\alpha_i}}(a_i|x_i)
\end{equation}
where $$ q_{\bar{\alpha_1}...\bar{ \alpha_{n+1}}}= \iiint_{\Lambda^{1...n}_{\bar{\alpha_1}...\bar{ \alpha_{n+1}}}} d\lambda_1...d\lambda_n\rho(\lambda_1,..., \lambda_n)\geq 0$$
and $\sum_{\bar{\alpha_1},...,\bar{ \alpha_{n+1}}}q_{\bar{\alpha_1}...\bar{ \alpha_{n+1}}}=1$. For instance, we consider here a simple system of four parties $A_i,\,(i=1,...,4)$. Eq.(\ref{N}) gives
 \begin{equation}\label{N1}
   P(a_1,a_2,a_3,a_4|x_1,x_2,x_3,x_4) = \sum_{\bar{\alpha_1},\bar{ \alpha_2},\bar{ \alpha_3},\bar{ \alpha_4}} q_{\bar{\alpha_1}\bar{ \alpha_2}\bar{ \alpha_3}\bar{ \alpha_4}}\Pi_{i=1}^4P_{\bar{\alpha_i}}(a_i|x_i)
\end{equation}
where $$ q_{\bar{\alpha_1}\bar{ \alpha_2}\bar{ \alpha_3}\bar{ \alpha_4}}= \int\int\int_{\Lambda^{123}_{\bar{\alpha_1}\bar{ \alpha_2}\bar{ \alpha_3}\bar{ \alpha_4}}} d\lambda_1d\lambda_2d\lambda_3\rho(\lambda_1,\lambda_2, \lambda_3)\geq 0.$$\\

Equation (\ref{N1}) represents the convex combination of deterministic strategies, the decomposition of local correlations,  where the weights $q_{\bar{\alpha_1}\bar{ \alpha_2} \bar{\alpha_3}\bar{\alpha_4}}$ represent the probability assigned by the sources to the strategies
$\bar{\alpha_1}, \bar{ \alpha_2},  \bar{\alpha_3}$ and $\bar{\alpha_4}$. ${\Lambda^{123}_{\bar{\alpha_1}\bar{ \alpha_2} \bar{\alpha_3}\bar{\alpha_4}}}$ represents all pairs $(\lambda_1, \lambda_2, \lambda_3)$ that specify the strategies $\bar{\alpha_i}$ for party $A_i$.  Now if we now consider $P$ as trilocal,  then the independence condition (\ref{M}) implies (for proof see Appendix A),
\begin{equation}\label{V}
q_{\bar{\alpha_1}\bar{\alpha_3}\bar{\alpha_4}} =  q_{\bar{\alpha_1}}q_{\bar{\alpha_3}\bar{\alpha_4}}\quad \forall \bar{\alpha_1}, \bar{\alpha_3},  \bar{\alpha_4}.
\end{equation}\\
\begin{equation}\label{W}
q_{\bar{\alpha_1}\bar{\alpha_2}\bar{\alpha_4}}  =  q_{\bar{\alpha_1}\bar{\alpha_2}} q_{\bar{\alpha_4}}\quad \forall  \bar{\alpha_1}, \bar{\alpha_2}, \bar{\alpha_4}.
\end{equation}
Any of these two equations (\ref{V}) and (\ref{W}) in turn implies
\begin{equation}\label{A4}
    q_{\bar{\alpha_1}\bar{\alpha_4}} = q_{\bar{\alpha_1}}q_{\bar{\alpha_4}}.
\end{equation}

\section{\textsc{Nonlinear Bell-type inequalities for $n$-local correlations}}
To study whether a given correlation is $n$-local or not,  we derive nonlinear Bell type inequalities which we refer as $\textit{$n$-local inequalities}.$ If a correlation violates  $\textit{$n$-local inequalities}$ then it is non-$n$-local in nature. In this context we consider two particular scenarios which occur frequently in various practical purposes:
\begin{center}
\begin{enumerate}
\item Scenario with binary inputs and outputs for each of the $n+1$ parties.
\item Scenario with binary inputs and outputs for each of the extreme two parties($A_1$ and $A_{n+1}$) and  with one input and four outputs for remaining $n-1$ parties.
\end{enumerate}
\end{center}
The second scenario is familiar with ideal entanglement swapping experiments where each of the intermediate parties perform full Bell basis measurement(one input and four outputs for each of the intermediate parties) thereby swapping entanglement from one extreme end of the network to another. But a complete Bell basis measurement is not always trivial enough to be executed. For instance(as already pointed out in \cite{BRAN}) in quantum linear optics(\cite{ver}) it is impossible to perform such an ideal joint measurement for the intermediate party in a network of three parties. Under such circumstances it will be interesting to consider the cases where each of the intermediate parties performs partial Bell basis measurements as the latter type of measurements is more feasible to perform than the former.  Hence from that experimental perspective it is interesting to deal with the first scenario.

\subsection{First Scenario}
Here we consider the case where each party $A_i$($i=1,...,n+1$) has binary inputs and outputs $x_i^k$ and $a_i^j$ respectively with $x_i^k,\,a_i^j\in\{0,1\}$. With $P^{22}$ denoting the conditional probability terms $P^{22}(a_1,...,a_{n+1}|x_1,...,x_{n+1})$, we define the $n$ partite correlation terms,
\begin{equation}\label{A8iii}
\langle A_{1,x_1},...A_{n+1,x_{n+1}}\rangle _{P^{22}}=\sum_{a_1,...,a_{n+1}}(-1)^{\sum_{i=1}^{n+1} a_i}P^{22}(a_1,...,a_{n+1}|x_1,...,x_{n+1})
\end{equation}
together with the terms $I^{22}_{A_2,...A_n}$ and $J^{22}_{A_2,...A_n}$(\cite{BRAN}) as
\begin{equation}\label{A8ii}
I^{22}_{A_2,...A_n}=\frac{1}{4}\sum_{x_1, x_{n+1}=0, 1}\langle A_{1,x_1},A_{2,0},...,A_{n,0},A_{n+1,x_{n+1}}\rangle _{P^{22}}.
\end{equation}

\begin{equation}\label{A9}
J^{22}_{A_2,...A_n}=\frac{1}{4}\sum_{x_1, x_{n+1}=0, 1}(-1)^{x_1+ x_{n+1}}\langle A_{1,x_1},A_{2,1},...,A_{n,1},A_{n+1,x_{n+1
}}\rangle _{P^{22}}.
\end{equation}\\
With the aid of these correlators we now frame the $n-$local inequality:

\textbf{Theorem 1}: \textsl{If $P^{22}$ is $n$-local then the following nonlinear Bell-type inequality holds}
\begin{equation}\label{A10}
\sqrt{\mid I^{22}_{A_2,...A_n}\mid}+  \sqrt{\mid J^{22}_{A_2,...A_n}\mid}\leq 1.
\end{equation}\\
For proof see Appendix(B). The above equations show that the $n$-local and the $\textit{bilocal}$ inequalities have the same form in the binary inputs and outputs scenario.\\

\subsubsection{Tightness of the $n$-local inequality}

The $n$-local inequality (\ref{A10}) is tight. To prove it we give an explicit $n$-local decomposition of correlations which satisfy Eq.(\ref{A10}). Let the correlation shared by $n+1$ parties be of the form:
\begin{center}
\begin{eqnarray*}
  P(a_1|x_1,\lambda_1,\eta_1) &=& 1,\,\, \textmd{if}\, a_1=\lambda_1\bigoplus\eta_1\ast x_1 \\
   &=& 0, \, \textmd{elsewhere}
\end{eqnarray*}
\begin{eqnarray*}
  P(a_{n+1}|x_{n+1},\lambda_n,\eta_2) &=& 1,\,\, \textmd{if}\,  a_{n+1}=\lambda_n\bigoplus\eta_2\ast x_{n+1} \\
   &=& 0, \, \textmd{elsewhere}
\end{eqnarray*}
\begin{eqnarray*}
  P(a_i|x_i,\lambda_{i-1},\lambda_i) &=& 1,\,\, \textmd{if}\, a_i=\lambda_{i-1}\bigoplus\lambda_i\\
   &=& 0, \, \textmd{elsewhere}\,\,\,\forall i=2,...,n
\end{eqnarray*}
\begin{eqnarray*}
  \rho_i(\lambda_i=0) &=& \frac{1}{2} \\
  \rho_i(\lambda_i=1) &=& \frac{1}{2},\,\,\,\forall i=1,...,n
\end{eqnarray*}
\begin{eqnarray*}
  \kappa_i(\eta_i=0) &=& r \\
  \kappa_i(\eta_i=1) &=& 1-r,\,\,\,i=1,2
\end{eqnarray*}
\end{center}
where each $\lambda_i$ is a random variable shared between two adjacent parties $A_i$ and $A_{i+1}$($i=1,...,n$) whereas  $\eta_i$ is the source of local randomness of party $A_i(i=1,n+1)$ and $r\in[0,1]$. Clearly this form of correlation gives $I^{22}_{A_2,...A_n}=r^2$ and $J^{22}_{A_2,...A_n}=(1-r)^2$. Hence Eq.(\ref{A10}) is satisfied.

\subsubsection{Local bounds}
\textbf{Theorem 2}: \textsl{If $P^{22}$ is local then it satisfies the inequality:}
\begin{equation}\label{A11}
\mid I^{22}_{A_2,...A_n}\mid +\mid J^{22}_{A_2,...A_n}\mid \leq 1.
\end{equation}\\
This can be proved in a pattern similar to that of the previous theorem except that here in place of Holder's inequality the inequality $ms +nt \leq (m+n)(s+t)$(for any $m$, $s$, $n$ and $t>0$)is to be used.

\begin{figure}[htb]
\centering
\includegraphics[width=3in]{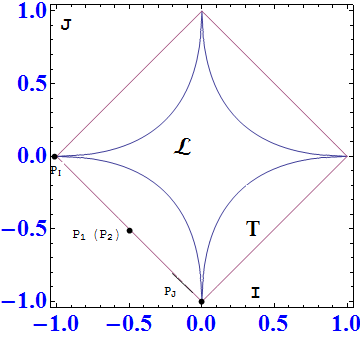}
\caption{\emph{We take projection of the $n+1$ partite correlation space in the $(I,\,J)$  plane where $I = I^{22}_{A_2,...A_n}(I^{14}_{A_2,...A_n})$ and $J = J^{22}_{A_2,...A_n}(J^{14}_{A_2,...A_n})$ as defined in equation (\ref{A8ii})[(\ref{A17})] and (\ref{A9})[(\ref{A18})]. $n$-local set $\mathcal{L}$ is bounded by the inequality $\sqrt{I} + \sqrt{J}\leq 1$. $n$-local set($\mathcal{T}$) is surrounded by local set, where local set satisfies the inequality $|I| + |J| =1$. The point $P_2(P_1)$ representing quantum correlation given by Eq.(\ref{A26})[(\ref{A26i})] lie outside the $n$-local polytope but lies on the facet of local polytope. $P_1(P_2)$ is the convex combination of the two extreme points $P_I(-1,0)$ and $P_J(0,-1)$.}}
\end{figure}
\subsection{Second Scenario}
In this case each party $A_i(i=2,...,n)$ have one input and four outputs while the extreme two parties, i.e., $A_1$ and $A_{n+1}$ both have two inputs and two outputs. The notations used here are similar as those introduced in (\cite{BRAN}) for $ P^{14}$ case, i.e., outputs of $A_{i}$ are denoted by a string of two bits $\textbf{a}_i=a_i^{0}a_i^{1}$ with $a_i^{0},\,a_i^{1}\in \{0,1\}$. It is different from the previous case where each bit $a_i^{j}(j=1,2)$  was output of $A_i$ for two different inputs ($x_i=0$ and $x_i=1$ respectively) for all $i\in\{2,...,n\}$.
\begin{equation}\label{A16}
  \langle A_{1,x_1}A_{2}^{x_2}....A_{n}^{x_n}A_{n+1,x_{n+1}}\rangle _{P^{14}}=\sum_{a_1,a_{n+1},a_2^0,a_2^1,...,a_n^0,a_n^1}(-1)^{a_1+a_{n+1}+\sum_{i=2}^n a_i^{x_i}}P^{14}(a_1, a_2^{0}a_2^{1},...a_n^0a_n^1,a_{n+1}|x_1,x_{n+1})
\end{equation}
together with the linear combination terms $I^{14}_{A_2,...A_n}$ and $J^{14}_{A_2,...A_n}$(\cite{BRAN}) as follows:
\begin{equation}\label{A17}
I^{14}_{A_2,...A_n}=\frac{1}{4}\sum_{x_1, x_{n+1}=0, 1}\langle  A_{1,x_1}A_{2}^{0}....A_{n}^{0}A_{n+1,x_{n+1}}\rangle _{P^{14}}.
\end{equation}

\begin{equation}\label{A18}
J^{14}_{A_2,...A_n}=\frac{1}{4}\sum_{x_1, x_{n+1}=0, 1}(-1)^{x_1+x_{n+1}}\langle  A_{1,x_1}A_{2}^{1}....A_{n}^{1}A_{n+1,x_{n+1}}\rangle _{P^{14}}.
\end{equation}
The correlators $\langle A_{1,x_1}A_{2}^{x_2}....A_{n}^{x_n}A_{n+1,x_{n+1}}\rangle_{P^{14}}$ given by Eq.(\ref{A16}) differ from the correlators $\langle A_{1,x_1},...A_{n+1,x_{n+1}}\rangle _{P^{22}}$(Eq.\ref{A8iii}). The difference mainly lies in the notation for the intermediate parties: in $P^{22}$ scenario $A_{i,x_i}\,(i=2,...,n)$ stands for different inputs($x_i\in\{0,1\}$) but for $P^{14}$ case, $A_{i}^{x_i}\,(i=2,...,n)$ mainly indicates which of the two bits of the corresponding output string is to be chosen.
As in the former case, $P^{14}$ correlations satisfy similar type of nonlinear inequality.
\\

\textbf{Theorem 3}: \textsl{If $P^{14}$ is $n$-local then the following nonlinear inequality necessarily holds,}

\begin{equation}\label{A19}
  \sqrt{\mid I^{14}_{A_2,...A_n}\mid}+  \sqrt{\mid J^{14}_{A_2,...A_n}\mid}\leq 1.
 \end{equation}\\

Proof: This inequality (\ref{A19}) can be derived directly from the inequality (\ref{A10}).

\begin{equation}
\begin{split}
P^{22}(a_1,...,a_{n+1}|x_1,...,x_{n+1})&=P^{14}(a_1, \overrightarrow{a_2}=a_2^{x_2},...,  \overrightarrow{a_n}=a_n^{x_n}, a_{n+1}|x_1,x_{n+1}) \\
  &=\sum_{a_2^0,a_2^1,...a_n^0,a_n^1}\delta_{\overrightarrow{a_2},a_2^{x_2}}....\delta_{\overrightarrow{a_n},a_2^{x_n}}P^{14}(a_1,a_2^0a_2^1,...,a_n^0a_n^1,a_{n+1}
  |x_1,x_{n+1}).
\end{split}
\end{equation}
This shows that from the correlation $P^{14}$, $n+1$ parties can obtain  a correlation $P^{22}$. Now from Eq.(\ref{A8iii}),

\begin{equation}\label{A8i}
\begin{split}
\langle A_{1,x_1},...A_{n+1,x_{n+1}}\rangle _{P^{22}}&=\sum_{a_1,...,a_{n+1}}(-1)^{\sum_{i=1}^{n+1} a_i}P^{22}(a_1,...,a_{n+1}|x_1,x_{n+1}) \\
       &=\sum_{a_1,...,a_{n+1}}(-1)^{\sum_{i=1}^{n+1} a_i}\sum_{a_2^0,a_2^1,...a_n^0,a_n^1}\delta_{\overrightarrow{a_2},a_2^{x_2}}....\delta_{\overrightarrow{a_n},a_2^{x_n}}P^{14}(a_1,a_2^0a_2^1,...,a_n^0a_n^1,a_{n+1}|x_1,x_{n+1})\\
       &=\sum_{a_1,a_2^0,a_2^1,...a_n^0,a_n^1,a_{n+1}}(-1)^{a_1+a_{n+1}+\sum_{i=2}^n a_i^{x_i}}P^{14}(a_1,a_2^0a_2^1,...,a_n^0a_n^1,a_{n+1}|x_1,x_{n+1})\\
       &= \langle A_{1,x_1}A_{2}^{x_2}....A_{n}^{x_n}A_{n+1,x_{n+1}}\rangle _{P^{14}}.
\end{split}
\end{equation}
Hence the values of $I^{22}_{A_2,...A_n}$ and $J^{22}_{A_2,...A_n}$ coincide with the values of $I^{14}_{A_2,...A_n}$ and $J^{14}_{A_2,...A_n}$ as defined by the equations (\ref{A17}) and (\ref{A18}) respectively. Let $P^{14}$ be $n$-local. As the process from $P^{14}$ to $P^{22}$ is made by $A_2,\,...,A_n$ locally,  therefore, $P^{22}$is also $n$-local and hence it satisfies equation (\ref{A10}). Thus, $I^{22}_{A_2,...A_n}=I^{14}_{A_2,...A_n}$ and $J^{22}_{A_2,...A_n}=J^{14}_{A_2,...A_n}$ jointly imply
the relation (\ref{A19}) is satisfied. $\blacksquare$\\

\subsubsection{Tightness of the $n$-local inequality}

 To prove tightness of $n$-local inequality Eq.(\ref{A19}) we proceed by considering the following correlation:
 \begin{center}
\begin{eqnarray*}
  P(a_1|x_1,\lambda_1,\eta_1) &=& 1,\,\, \textmd{if}\, a_1=\lambda_1\bigoplus\eta_1\ast x_1 \\
   &=& 0 \, \textmd{elsewhere}
\end{eqnarray*}
\begin{eqnarray*}
  P(a_{n+1}|x_{n+1},\lambda_n,\eta_2) &=& 1,\,\, \textmd{if}\,  a_{n+1}=\lambda_n\bigoplus\eta_2\ast x_{n+1} \\
   &=& 0 \, \textmd{elsewhere}
\end{eqnarray*}
\begin{eqnarray*}
  P(a_i^0a_i^1|x_i,\lambda_{i-1},\lambda_i) &=& 1,\,\, \textmd{if}\, a_i^0\ast a_i^1=\lambda_{i-1}\bigoplus\lambda_i\\
   &=& 0 \, \textmd{elsewhere}\,\,\,\forall i=2,...,n
\end{eqnarray*}
\begin{eqnarray*}
  \rho_i(\lambda_i=0) &=& \frac{1}{2} \\
  \rho_i(\lambda_i=1) &=& \frac{1}{2},\,\,\,\forall i=1,...,n
\end{eqnarray*}
\begin{eqnarray*}
  \kappa_i(\eta_i=0) &=& r \\
  \kappa_i(\eta_i=1) &=& 1-r,\,\,\,\forall i=1,2
\end{eqnarray*}
\end{center}
where $\eta_1$, $\eta_2$, $\lambda_i(i=1,...,n)$ and $r$ have the same terminology as in the previous scenario. Clearly this form of correlation gives $I^{14}_{A_2,...A_n}=r^2$ and $J^{14}_{A_2,...A_n}=(1-r)^2$. Hence Eq.(\ref{A19}) is satisfied.

\subsubsection{Local Bounds}

As $P^{22}$ can be obtained from $P^{14}$ by changing the number of inputs and outputs for $A_i(i=2,...,n)$ locally,  therefore, if $P^{14}$ is local, then it satisfies the non-linear inequality of the same form as in the $P^{22}$ case;
\begin{equation}\label{A21}
  \mid I^{14}_{A_2,...A_n}\mid+ \mid J^{14}_{A_2,...A_n}\mid\leq 1.
\end{equation}
\section{Quantum Correlations in  $n$-local Scenario}
Here we discuss about the quantum correlations in $P^{22}$ and $P^{14}$ scenarios. While $P^{14}$ scenario is familiar with ideal entanglement swapping experiment, $P^{22}$ scenario is associated with the experiment where each of the $n$ intermediate parties can partially distinguish between the Bell states. In both the scenarios quantum correlations violate \textit{$n$-local} bounds (Eqs.(\ref{A10},\ref{A19})).
\begin{center}
\begin{itemize}
\item In $P^{14}$,  entanglement swapping experiment with full Bell state measurement.
\item In $P^{22}$, we perform partial Bell-state measurement.
\end{itemize}
\end{center}
In general, for both the cases  we consider a model where each of $n$ independent quantum sources $S_{i}(i=1,...,n)$  sends a particle to parties $A_i$ and $A_{i+1}$ in the state $\varrho_{i}$ so that the  overall quantum state is
\begin{equation}\label{11}
  \varrho_{A_1,...A_{n}}=\otimes_{i=1}^{n}\varrho_{i}.
\end{equation}

\subsection{Entanglement Swapping with a complete Bell-State measurement}
In this case, $A_1$ and $A_{n+1}$ are supposed to have binary inputs and binary outputs whereas remaining $n-1$ parties $A_i(i=2,...,n)$ will perform a full Bell basis measurement (each of them have one input and four outputs). Besides we assume that each source $S_i$ produces the same Bell state $|\psi^{-}\rangle$.
In this scenario, party $A_i(i=2,...,n)$ receives part of the state $\varrho_{i-1}$ from $S_{i-1}$ and part of the state $\varrho_{i+1}$ from $S_{i+1}$. It then performs complete Bell-state measurement on its two particles; the four possible outcomes $\textbf{a}_i=a_i^{0}a_i^{1}=00, 01, 10, 11$ that can be obtained correspond to the four Bell states (with standard notations) $|\phi^{+}\rangle, |\phi^{-}\rangle, |\psi^{+}\rangle $ and $|\psi^{-}\rangle$ respectively. The resulting state is a bipartite entangled state shared by $A_{i-1}$ and $A_{i+1}$. Let it be denoted by $\varrho_{i-1,i+1}$. Ultimately parties $A_1$ and $A_{n+1}$ share an entangled state $\varrho_{1,n+1}$. Now, $A_1$ and $A_{n+1}$ perform the following measurements(\cite{BRAN}):
\begin{equation}\label{m}
\begin{split}
\hat{A}_1^{0}&=\hat{A}_{n+1}^{0}=\frac{\sigma_{z}+\sigma_{x}}{\sqrt{2}} (\textmd{for}\, x_1=x_{n+1}=0)\, \textmd{or}\\
\hat{A}_{1}^{1}&=\hat{A}_{n+1}^{1}=\frac{\sigma_{z}-\sigma_{x}}{\sqrt{2}} (\textmd{for}\,  x_1=x_{n+1}=1).
\end{split}
\end{equation}
The correlations are of the form:
\begin{equation}\label{A26}
P^{14}(a_1, a_2^{0}a_2^{1},...a_n^0a_n^1,a_{n+1}|x_1,x_{n+1})_{Q}=\frac{1+(-1)^{a_1+a_{n+1}+1}(\frac{(-1)^{\sum_{i=2}^{n}a_i^0}+(-1)^{\sum_{i=2}^{n}a_i^1+x_1+x_{n+1}}}{2})}{2^{2n}}.
\end{equation}

From the definitions (\ref{A17}) and (\ref{A18}) it can be checked that :
\begin{equation}\label{24}
I^{14}_{A_2,...A_n}(P_{Q}^{14})= -\frac{1}{2}; \quad J^{14}_{A_2,...A_n}(P_{Q}^{14})=-\frac{1}{2}
\end{equation}
Hence the above correlation shows quantum violation of $n$-locality (Eq.(\ref{A19})) but satisfies the locality constraint (Eq.(\ref{A21})). However, $P_{Q}^{14}$ can be obtained as a convex combination two correlations: $P_{Q}^{14}=\frac{P_{I}^{14}+P_J^{14}}{2}$ where:
\begin{equation}\label{s1}
\begin{split}
P_I^{14}(a_1, a_2^{0}a_2^{1},...,a_n^0a_n^1,a_{n+1}|x_1,x_{n+1})_{Q}&=\frac{[1+(-1)^{a_1+a_{n+1}+1}(\frac{(-1)^{\sum_{i=2}^{n}a_i^0}}{2})]}{2^{2n-1}}\\
P_J^{14}(a_1, a_2^{0}a_2^{1},...,a_n^0a_n^1,a_{n+1}|x_1,x_{n+1})_{Q}&=\frac{(1+(-1)^{a_1+a_{n+1}+1}(\frac{(-1)^{\sum_{i=2}^{n}a_i^1+x_1+x_{n+1}}}{2}))}{2^{2n-1}}.
\end{split}
\end{equation}
For the correlation $P_I^{14}$, $I^{14}_{A_2,...A_n}=-1$, $J^{14}_{A_2,...A_n}=0$ whereas for $P_J^{14}$, we get $I^{14}_{A_2,...A_n}=0$ and $J^{14}_{A_2,...A_n}=-1.$ Hence $P_Q^{14}$ represents a point ($P_2$) on the facet of local polytope but lies outside the $n$-local polytope (FIG.4).

\subsection{Partial Bell-state measurement}
As in the previous case here each of the $n$ sources $S_i(i=1,...,n)$ sends Bell state $|\psi^-\rangle$.
$A_1$ and $A_{n+1}$ are supposed to have binary inputs and binary outputs, the measurements being the same as in the previous scenario, (Eq.\ref{m}) and remaining $n-1$ parties $A_i(i=2,...,n)$ will perform partial Bell-state measurement, i.e., $A_i(i=2,...,n)$ measures either:
\begin{equation}\label{m1}
\begin{split}
\hat{A}_i^{0}&=|\phi^{+}\rangle\langle\phi^{+}|+ |\phi^{-}\rangle\langle\phi^{-}|-|\psi^{+}\rangle\langle\psi^{+}|-|\psi^{-}\rangle\langle\psi^{-}|=\sigma_{z}\otimes\sigma_{z}\,\textmd{or}\\
\hat{A}_i^{1}&=|\phi^{+}\rangle\langle\phi^{+}|- |\phi^{-}\rangle\langle\phi^{-}|+|\psi^{+}\rangle\langle\psi^{+}|-|\psi^{-}\rangle\langle\psi^{-}|=\sigma_{x}\otimes\sigma_{x}.    \end{split}
\end{equation}
Each of the $n-1$ intermediate parties $A_i(i=2,...,n)$ receives part of the state $\varrho_{i-1}$ from $S_{i-1}$ and part of $\varrho_{i+1}$ from $S_{i+1}$ which then applies any one of the separable measurements (Eq.(\ref{m1})) on its two particles. The post measurement state is a mixed bipartite entangled state shared by $A_{i-1}$ and $A_{i+1}$. Let it be denoted by $\varrho_{i-1,i+1}$. Ultimately parties $A_1$ and $A_{n+1}$ shares an mixed entangled state $\varrho_{1,n+1}$. Finally, $A_1$ and $A_{n+1}$ perform local measurements Eq.(\ref{m}) on their respective part of states.
The correlations take the form:
\begin{equation}\label{A26i}
P^{22}(a_1,...,a_{n+1}|x_1,...,x_{n+1})= \frac{1+(-1)^{(\sum_{i=2}^n a_i+1)}\frac{\Pi_{i=2}^n\delta_{(x_i,0)}+(-1)^{x_1+x_{n+1}}\Pi_{i=2}^n\delta_{(x_i,1)}}{2}}{2^{n+1}}
\end{equation}
From the definitions (\ref{A8ii}) and (\ref{A9}) it can be checked that :
\begin{equation}\label{24}
I^{22}_{A_2,...A_n}(P_{Q}^{22})= -\frac{1}{2}; \quad J^{22}_{A_2,...A_n}(P_{Q}^{22})=-\frac{1}{2}
\end{equation}
Hence quantum violation of $n$-locality (Eq.(\ref{A10})) is obtained, but the correlation is local in nature as it satisfies Eq.(\ref{A11}). As in the previous scenario, $P_{Q}^{22}$ can be obtained from $P_{Q}^{22}=\frac{P_{I}^{22}+P_J^{22}}{2}$ where:
\begin{equation}\label{s1}
\begin{split}
P_I^{22}(a_1,...,a_{n+1}|x_1,...,x_{n+1})&= \frac{1+(-1)^{(\sum_{i=2}^n a_i+1)}\frac{\Pi_{i=2}^n\delta_{(x_i,0)}}{2}}{2^{n}}\\
P_J^{22}(a_1,...,a_{n+1}|x_1,...,x_{n+1})&= \frac{1+(-1)^{(\sum_{i=2}^n a_i+1)}\frac{(-1)^{x_1+x_{n+1}}\Pi_{i=2}^n\delta_{(x_i,1)}}{2}}{2^{n}}
\end{split}
\end{equation}
For $P_I^{22}$, $I^{22}_{A_2,...A_n}=-1$, $J^{22}_{A_2,...A_n}=0$ and for $P_J^{22}$, we get $I^{22}_{A_2,...A_n}=0$ and $J^{22}_{A_2,...A_n}=-1.$ Hence $P_Q^{22}$ represents a point($P_1$) on the facet of local polytope (FIG.4) lying outside the $n$-local polytope.

\subsubsection{Advantage of $n$-locality assumption in a network}
\textit{Resistance to noise:} As already pointed out by Branciard et.al. \cite{BRAN},  quantifying resistance to noise of the nonbilocal correlations is one way to realize the advantage of $n$-locality assumption in a network. For that let us consider  an entanglement swapping scenario where each of $n$ independent sources $S_i(i=1,...,n)$ produces a noisy two qubit state:
\begin{equation}\label{x4iii}
\chi_i=\alpha_i |\psi^-\rangle\langle\psi^-|+(1-\alpha_i)\frac{\mathbf{1}}{4},\,(\textmd{where}\,\alpha_i\in[0,1](i=1,...,n)).
\end{equation}
Here $\alpha_i$ is the visibility of $\chi_i(i=1,,...,n)$ which is a measure of the resistance to noise given by the state and is referred to as \textit{local visibility threshold} ($V$ \cite{BRAN}), i.e., the largest visibility ( $\alpha_i$) for which $\chi_i$ is local in usual CHSH sense \cite{Cla}. Analogously the largest visibility for which the correlation generated is $n$-local  is called the \textit{$n$-local visibility threshold} ($V_{nloc}=\prod_{i=1}^{n}\alpha_i$ \cite{BRAN}). Clearly, quantum advantage (i.e., demonstration of nonlocality apart from standard CHSH sense ) is obtained if $V_{nloc}<V$. With the measurement settings considered here for the $n+1$ partite system, for both $P^{14}$ and $P^{22}$ scenarios, $V_{nloc}=\frac{1}{2}$ which is same as in bilocal scenario \cite{BRAN}. For different measurement settings (which gives maximal violation of Bell-CHSH operator \cite{Cla}), $V=\frac{1}{\sqrt{2}}.$ Thus, $V_{nloc}<V$ and hence advantage is obtained when $n$locality condition is assumed in a network but the resistance to noise and hence the advantage cannot be further increased compared to bilocal scenario if number of parties is increased in a linear pattern unlike that in a star configuration (\cite{Tav}) where $V_{nloc}=\frac{1}{2^{\frac{n}{2}}}$. which increases as number of parties increases. This in turn gives rise to the intuition that mere increase in the number of intermediate parties cannot be useful in this context.

\section{Conclusion}
In recent times, models having independent systems (characterized by uncorrelated hidden states) are used to have a better insight regarding the nonlocal correlations simulated in many experiments based on entanglement swapping (which creates correlation between initially uncorrelated parties). In this respect \textit{bilocal} models (source independence in three party scenario) and other related topics were discussed in (\cite{BRAN}). Motivated by the bilocal scenario, emphasizing on the significance of source independence for practical demonstrations, we have tried to enhance the study of correlations characterized by independent sources thereby reviewing the topic of source independence in generalized $n+1$ party scenario. Clearly the nonlinear inequalities obtained in the $n$-local scenario maintains the same structure as that in $bilocal $ scenario. However the measurements considered here do not suffice to decrease the $n$-local visibility threshold($V_{nloc}$) compared to $V_{biloc}$(\cite{BRAN}). Even the possible change of measurement bases of the parties is of no use in this regard which in turn give rise to the intuition that increase in number of parties arranged in a linear pattern in a source independent network reduces to trivial party extension where one may assume the intermediate $n-1$ parties to behave like a  single party. Perhaps more generalized measurements settings specifically positive operator valued measurements (POVM(\cite{won1},\cite{won2})) may help to increase the resistance to noise in this type of linear network. One may try to modify and hence develop  new Bell-type inequalities compatible with $n$-local scenario. It may also be interesting to investigate further to develop any other pattern of arrangement of the parties which in turn may ensure increase of resistance to noise in the corresponding network characterized by source independence.   \\
 \\
{\bf Acknowledgement.} The authors are grateful to D. Rosset and C. Branciard for stimulating discussions on the topic while visiting Kolkata. The authors also thank Ajoy Sen for interesting and helpful discussions relating to the topic of this work. The author KM acknowledges the financial support by UGC, New Delhi.

{ \textit{Note Added:}} While the present work was under review we became aware of the work related to this topic\cite{Tav}.

\appendix{}
\section{Representation of the $4$-local correlation in terms of $q_{\bar{\alpha_1}\bar{ \alpha_2} \bar{\alpha_3}\bar{\alpha_4}}$}
We note here that $\bar{\alpha_1}$ is specified  by $\lambda_1;$ $ \bar{ \alpha_2}$ is specified by $\lambda_1, \lambda_2$;  $ \bar{\alpha_3}$ is specified by $\lambda_2, \lambda_3$; and  $ \bar{ \alpha_4}$ is specified by $\lambda_3$. Then $\bigcup_{\bar{ \alpha_2}}\Lambda^{123}_{\bar{\alpha_1}\bar{ \alpha_2} \bar{\alpha_3}\bar{\alpha_4}} = \Lambda^{123}_{\bar{\alpha_1}\bar{\alpha_3}\bar{\alpha_4}} = \Lambda^{1}_{\bar{\alpha_1}} \times {\Lambda^{23}_{\bar{\alpha_3}\bar{\alpha_4}}}$ ,  $\bigcup_{\bar{ \alpha_3}}\Lambda^{123}_{\bar{\alpha_1}\bar{ \alpha_2} \bar{\alpha_3}\bar{\alpha_4}} = \Lambda^{123}_{\bar{\alpha_1}\bar{\alpha_2}\bar{\alpha_4}} = \Lambda^{12}_{\bar{\alpha_1}\bar{\alpha_2}} \times {\Lambda^{3}_{\bar{\alpha_4}}}$ and\\

\begin{equation}\label{O}
    q_{\bar{\alpha_1}\bar{\alpha_3}\bar{\alpha_4}} =\sum_{\bar{\alpha_2}}q_{\bar{\alpha_1}\bar{ \alpha_2} \bar{\alpha_3}\bar{\alpha_4}}= \iiint_{\Lambda^{1}_{\bar{\alpha_1}} \times {\Lambda^{23}_{\bar{\alpha_3}\bar{\alpha_4}}}} d\lambda_1d\lambda_2d\lambda_3\rho(\lambda_1, \lambda_2, \lambda_3).
\end{equation}\\
\begin{equation}\label{P}
    q_{\bar{\alpha_1}\bar{\alpha_2}\bar{\alpha_4}} = \sum_{\bar{\alpha_3}}q_{\bar{\alpha_1}\bar{ \alpha_2} \bar{\alpha_3}\bar{\alpha_4}}= \iiint_{\Lambda^{12}_{\bar{\alpha_1}\bar{\alpha_2}} \times {\Lambda^{3}_{\bar{\alpha_4}}}} d\lambda_1d\lambda_2d\lambda_3\rho(\lambda_1, \lambda_2, \lambda_3).
\end{equation}\\
\begin{equation}\label{Q}
    q_{\bar{\alpha_1}\bar{\alpha_4}} = \sum_{\bar{\alpha_3}}q_{\bar{\alpha_1}\bar{\alpha_3}\bar{\alpha_4}} =  \iiint_{\Lambda^{1}_{\bar{\alpha_1}} \times {\Lambda^{2} \times \Lambda^{3}_{\bar{\alpha_4}}}} d\lambda_1d\lambda_2d\lambda_3\rho(\lambda_1, \lambda_2, \lambda_3).
\end{equation}\\
\begin{equation}\label{R}
    q_{\bar{\alpha_1}\bar{\alpha_2}} = \sum_{\bar{\alpha_4}}q_{\bar{\alpha_1}\bar{\alpha_2}\bar{\alpha_4}} =   \iiint_{\Lambda^{12}_{\bar{\alpha_1}\bar{\alpha_2}} \times {\Lambda^{3}}} d\lambda_1d\lambda_2d\lambda_3\rho(\lambda_1, \lambda_2, \lambda_3).
\end{equation}\\
\begin{equation}\label{S}
q_{\bar{\alpha_3}\bar{\alpha_4}} = \sum_{\bar{\alpha_1}}q_{\bar{\alpha_1} \bar{\alpha_3}\bar{\alpha_4}} = \iiint_{\Lambda^{1} \times {\Lambda^{23}_{\bar{\alpha_3}\bar{\alpha_4}}}} d\lambda_1d\lambda_2d\lambda_3\rho(\lambda_1, \lambda_2, \lambda_3).
\end{equation}\\
\begin{equation}\label{T}
q_{\bar{\alpha_1}} = \sum_{\bar{\alpha_4}}q_{\bar{\alpha_1}\bar{\alpha_4}} = \iiint_{\Lambda^{1}_{\bar{\alpha_1}} \times \Lambda^{2} \times \Lambda^{3}} d\lambda_1d\lambda_2d\lambda_3\rho(\lambda_1, \lambda_2, \lambda_3).
\end{equation}\\
\begin{equation}\label{U}
q_{\bar{\alpha_4}} = \sum_{\bar{\alpha_1}}q_{\bar{\alpha_1}\bar{\alpha_4}} = \iiint_{\Lambda^{1} \times \Lambda^{2} \times \Lambda^{3}_{\bar{\alpha_4}}} d\lambda_1d\lambda_2d\lambda_3\rho(\lambda_1, \lambda_2, \lambda_3).
\end{equation}\\
where $\Lambda^1 = \bigcup_{\bar{\alpha_1}}\Lambda^{1}_{\bar{\alpha_1}}$ ,  $\Lambda^2 =\bigcup_{\bar{\alpha_2}\bar{\alpha_3}}\Lambda^{2}_{\bar{\alpha_2}\bar{\alpha_3}}$ and $\Lambda^3 = \bigcup_{\bar{\alpha_4}}\Lambda^{3}_{\bar{\alpha_4}}$ are the corresponding state spaces of the variables $\lambda_1, \lambda_2$ and $\lambda_3$ respectively.\\
 Now if we now consider $P$ as $4$-local,  then the independence condition (\ref{M}) implies (equation (\ref{V}) is obtained from (\ref{O}), (\ref{S}), (\ref{T}) and equation (\ref{W}) is obtained from (\ref{P}), (\ref{R}), (\ref{U})) for all $\bar{\alpha_1}, \bar{\alpha_3}$ and $\bar{\alpha_4}$,
\begin{equation}\label{V}
q_{\bar{\alpha_1}\bar{\alpha_3}\bar{\alpha_4}} =  q_{\bar{\alpha_1}}q_{\bar{\alpha_3}\bar{\alpha_4}}\quad \forall \bar{\alpha_1}, \bar{\alpha_3},  \bar{\alpha_4}.
\end{equation}\\
\begin{equation}\label{W}
q_{\bar{\alpha_1}\bar{\alpha_2}\bar{\alpha_4}}  =  q_{\bar{\alpha_1}\bar{\alpha_2}} q_{\bar{\alpha_4}}\quad \forall  \bar{\alpha_1}, \bar{\alpha_2}, \bar{\alpha_4}.
\end{equation}

The above result can be easily extended to $n$-local scenario.
\section{Proof of Eq.(\ref{A10})}
 We define,
\begin{eqnarray}
\langle A_{1,x_1}\rangle _{\lambda_{1}}&=&\sum_{a_1}(-1)^{a_1}P^{22}(a_1|x_1,\lambda_{1})\\
\langle A_{i,x_i}\rangle_{\lambda_{i}}&=&\sum_{a_i}(-1)^{a_i}P^{22}(a_i|x_i,\lambda_{i},\lambda_{i+1}),\,i=2,...,n\\
\langle A_{n+1,x_{n+1}}\rangle _{\lambda_{n}}&=&\sum_{a_{n+1}}(-1)^{a_{n+1}}P^{22}(a_{n+1}|x_{n+1},\lambda_{n}).
\end{eqnarray}
Since by assumption $P^{22}$  is $n$-local, it has a $n$-local decomposition of the form (\ref{L}) and (\ref{M}). So we get,
$$
I^{22}_{A_2,...A_n}=\\
\frac{1}{4}\int\int...\int d\lambda_{1},... d\lambda_{n} \Pi_{i=1}^n\rho_{i}(\lambda_{i})(\langle A_{1,0}+A_{1,1}\rangle _{\lambda_{1}})(\langle A_{n+1,0}+A_{n+1,1}\rangle _{\lambda_{n}})(\langle A_{2,0}....A_{n-1,0}\rangle)_{\lambda_{1}....\lambda_{n}}.$$
Now,
\begin{equation}
\mid \langle A_{2,0}....A_{n-1,0}\rangle_{\lambda_{1}....\lambda_{n}} \mid \leq 1.
\end{equation}
Using the above relation, we have,
\begin{equation*}
\begin{split}
\mid I^{22}_{A_2,...A_n}\mid  &\leq \frac{1}{4}\int\int d\lambda_{1} d\lambda_{n}\rho_{1}(\lambda_{1})\rho_{n}(\lambda_{n}) (\langle A_{1,0}+A_{1,1}\rangle _{\lambda_{1}})(\langle A_{n+1,0}+A_{n+1,1}\rangle _{\lambda_{n}})\int....\int\Pi_{i=2}^{n-1}\rho_{i}(\lambda_{i})\\
&=  \int d\lambda_{1}\rho_{1}(\lambda_{1}) \frac{\mid(\langle A_{1,0}+A_{1,1}\rangle )_{\lambda_{1}}\mid}{2}\times\int d\lambda_{n}\rho_n(\lambda_{n}) \frac{\mid(\langle A_{n+1,0}+A_{n+1,1}\rangle )_{\lambda_{n}}\mid}{2}.
\end{split}
\end{equation*}
Similarly for $J^{22}_{A_2,...A_n}$ it can be shown that,

\begin{equation*}
\mid J^{22}_{A_2,...A_n}\mid  \leq\int d\lambda_{1}\rho_{1}(\lambda_{1}) \frac{\mid(\langle A_{1,0}-A_{1,1}\rangle )_{\lambda_{1}}\mid}{2}\times\int d\lambda_{n}\rho_n(\lambda_{n}) \frac{\mid(\langle A_{n+1,0}-A_{n+1,1}\rangle )_{\lambda_{n}}\mid}{2}.
\end{equation*}
Now,  by using Holder's inequality for $4$ positive quantities $\mid(\langle A_{1,0}+A_{1,1}\rangle )_{\lambda_{1}}\mid$, $\mid(\langle A_{1,0}-A_{1,1}\rangle )_{\lambda_{1}}\mid$, $\mid(\langle A_{n+1,0}+A_{n+1,1}\rangle )_{\lambda_{n}}\mid$, $\mid(\langle A_{n+1,0}-A_{n+1,1}\rangle )_{\lambda_{n}}\mid$ we get,
\begin{equation*}
\begin{split}
\sqrt{\mid I^{22}_{A_2,...A_n}\mid}+  \sqrt{\mid J^{22}_{A_2,...A_n}\mid}&\leq \sqrt{\int d\lambda_{1}\rho_{1}(\lambda_{1})(\frac{\mid(\langle A_{1,0}+A_{1,1}\rangle )_{\lambda_{1}}\mid}{2}+\frac{\mid(\langle A_{1,0}-A_{1,1}\rangle )_{\lambda_{1}}\mid}{2})}\\
 & \leq \sqrt{\int d\lambda_{n}\rho_{n}(\lambda_{n})(\frac{\mid(\langle A_{n+1,0}+A_{n+1,1}\rangle )_{\lambda_{n}}\mid}{2}+\frac{\mid(\langle A_{n+1,0}-A_{n+1,1}\rangle )_{\lambda_{n}}\mid}{2})}.
\end{split}
\end{equation*}
Again,  $\frac{\mid(\langle A_{1,0}+A_{1,1}\rangle )_{\lambda_{1}}\mid}{2}+\frac{\mid(\langle A_{1,0}-A_{1,1}\rangle )_{\lambda_{1}}\mid}{2}$ = $\max(\mid \langle A_{1,0}\rangle _{\lambda_{1}}\mid, \mid \langle A_{1,1}\rangle _{\lambda_{1}}\mid)\leq 1$ and similarly $\max(\mid \langle A_{n+1,0}\rangle _{\lambda_{1}}\mid, \mid \langle A_{n+1,1}\rangle _{\lambda_{1}}\mid)\leq 1$. Using these we get,

\begin{equation}
\begin{split}
\sqrt{\mid I^{22}_{A_2,...A_n}\mid}+  \sqrt{\mid J^{22}_{A_2,...A_n}\mid} & \leq \int d\lambda_{1}\rho_1(\lambda_{1}).\int d\lambda_{n}\rho_n(\lambda_{n}) \\
& = 1.
\end{split}
\end{equation}
Hence the inequality (\ref{A10}) is satisfied. $\blacksquare$\\
\end{document}